\documentclass[preprint2]{aastex}

\shorttitle{The Baby Homunculus}
\shortauthors{Z. Abraham, D. Falceta=Gon\c calves, P.P.B. Beaklini}

\begin{document}

\title{$\eta$ Carinae Baby Homunculus Uncovered by ALMA}

\author{Zulema Abraham}
\affil{Instituto de Astronomia, Geof\'isica e Ci\^encias Atmosf\'ericas, 
Universidade de S\~ao Paulo, R. do Mat\~ao 1226,\\ Cidade Universit\'aria, CEP 05508-900, S\~ao Paulo, SP,
 Brazil}
\email{zulema.abraham@iag.usp.br} 
	
\author{Diego Falceta-Gon\c calves\altaffilmark{1}}
\affil{Escola de Artes, Ci\^encias e Humanidades, Universidade de S\~ao Paulo, R. Arlindo Bettio 1000, 03828-000, S\~ao Paulo, SP, Brazil}
\altaffiltext{1}
{SUPA, School of Physics \& Astronomy, University of St Andrews, North Haugh, St Andrews, Fife KY169SS, UK}

\and

\author{Pedro P.B. Beaklini\altaffilmark{2}}
\affil{Instituto de Astronomia, Geof\'isica e Ci\^encias Atmosf\'ericas, 
Universidade de S\~ao Paulo, R. do Mat\~ao 1226,\\ Cidade Universit\'aria, CEP 05508-900, S\~ao Paulo, SP,
 Brazil}
\altaffiltext{2}
{IRAM, Institut de RadioAstronomie Millim\'etrique,300 rue de la Piscine, Domaine Universitaire 38406 Saint Martin d'H\`eres, France
}

\begin{abstract}
We report observations  of $\eta$ Carinae obtained with ALMA in the continuum of 100, 230, 280 and 660 GHz in 2012 November,  with a resolution that varied from  $2\farcs88$ to $0\farcs45$ for the lower and higher frequencies respectively.  The source is not resolved, even at the highest frequency;  its  spectrum  is characteristic of thermal bremsstrahlung of a compact source,  but different from the spectrum of optically thin wind. The recombination lines H42$\alpha$, He42$\alpha$, H40$\alpha$, He40$\alpha$, H50$\beta$, H28$\alpha$, He28$\alpha$, H21$\alpha$ and He21$\alpha$ were also detected  and their intensities  reveal  non local thermodynamic equilibrium (NLTE) effects.  We found that the line profiles could only be fit by an expanding shell of dense and ionized gas, which produces a slow shock in the surroundings of $\eta$ Carinae. Combined with fittings to the continuum, we were able to constrain the shell size, radius, density, temperature and velocity.  The detection of the He recombination lines is compatible with the high temperature gas and requires a high energy ionizing photon flux, which must be provided by the companion star.  The  mass loss rate and wind velocity, necessary to explain the formation of the shell, are compatible with a LBV eruption. The position, velocity and physical parameters of the shell coincide with those of the Weigelt blobs.  The dynamics found for the expanding shell corresponds to matter ejected by $\eta$ Carinae in 1941, in an event similar to that which  formed the Little Homunculus ; for that reason we called the new ejecta the "Baby Homunculus".

\end{abstract}


\section{Introduction}
The Homunculus Nebula, named by \citet{gav50}, is a bipolar shell of about 12 M$_\odot$  of gas and dust \citep{smi03}, ejected from the massive star $\eta$ Carinae during the $Great$ $Eruption$, in the decade of 1840. As the nebula expanded with an average velocity of 650 km s$^{-1}$ to its actual size of about  18$\arcsec$ or 0.089 pc (assuming a distance to $\eta$ Carinae of 2.3 kpc), it cooled down forming a thick dust layer that absorbed the stellar light \citep{hum99,smi98}, slowly  decreasing its luminosity and  increasing the visual magnitude of $\eta$ Carinae from -1 in 1843 to 7.4 in 1887. In that year, a new and sudden increase in the stellar luminosity was detected, which lasted for seven years and was interpreted as a new but minor episode of mass ejection, dimmed by dust absorption. The 0.1 M$_\odot$ of ejected matter remained inside the Homunculus Nebula and it is now at a distance of about $\pm 2\arcsec$ from the star,  expanding at a velocity of about 200 km s$^{-1}$; it was  detected  through long-slit mapping of several  [\ion{Fe}{2}] lines, using the  Space Telescope 
Imaging Spectrograph and named the  Little Homunculus (LH) by \citet{ish03}. 

The LH was also identified in the  8.6 GHz (3 cm) and 5.4 GHz (6 cm) radio continuum and hydrogen  (H91$\alpha$ and H106$\alpha$) recombination line maps , obtained with the Australian Compact Array with $1\arcsec$ resolution \citep{dun97,smi05,teo08}. On the other hand, observations at 230 GHz (1 mm) and 100 GHz (3 mm), obtained with the SEST radiotelescope with resolution of $30\arcsec$ and $50\arcsec$ respectively, showed a much stronger unresolved radio source at the star position, which was interpreted as free-free emission from an optically thick wind of spectral index 1.3 \citep{cox95a}. Recombination line observations (H40$\alpha$, H30$\alpha$, H29$\alpha$, H50$\beta$, and H37$\beta$) at these wavelength bands, showed strong emission at the compact source position, with evidences of departure from local thermodynamic equilibrium (NLTE),  implying gas densities of more than  $10^7$ cm$^{-3}$ and electron temperatures of 15,000 K \citep{cox95b}.

 The origin of the eruptions in $\eta$ Carinae is still under discussion, it includes LBV behavior \citep{hum79}, supernova impostors \citep{smi11a} or violent star collisions  \citep{smi11}. This last assumption is based on the binary nature of the $\eta$ Carinae system, revealed by the strict 5.54 y periodicity of the minima in the high excitation line emission, X-ray and radio continuum light curves \citep{dam96,cor01,abr05}. The secondary star, in a highly eccentric orbit, has not been observed directly, but a range of possible surface temperatures, higher than that of $\eta$ Carinae, was determined through photoionization models \citep{meh10} of the high excitation lines in the Weigelt blobs \citep{wei86}.

In this paper we report continuum and, \ion{H}{1} and \ion{He}{1} recombination line observations (42$\alpha$, 40$\alpha$, 30$\alpha$, 28$\alpha$, 21$\alpha$ and 50$\beta$) obtained with ALMA between 100 and 660 GHz (3 mm  and 455 $\micron$, respectively), with resolutions ranging between $2\farcs88$ and $0\farcs45$.   These are the first  observations of $\eta$ Carinae at such high frequencies and spatial resolutions. The observations show that the continuum spectrum has a turnover between 100 and 230 GHz, indicating that it arises in a compact \ion{H}{2} region instead of an ionized wind.  The recombination line intensities confirm the departure from LTE and the electron densities derived by \citet{cox95b} and constrain the geometry of the emitting region. In addition, the spectra show for the first time, besides of the H lines, the presence of the corresponding \ion{He}{1} lines. 

In Section 2 we describe the observations, in Section 3 the continuum and recombination line maps, as well as the total spectrum of the region. In Section 4 we discuss the continuum spectrum  and recombination line intensities and profiles and obtained the parameters of the emitting region, modeled as a spherical shell. Finally in Section 5 we summarize our conclusions. 

\begin{figure*}[!t]
\begin{center}
\includegraphics[width= 15cm]{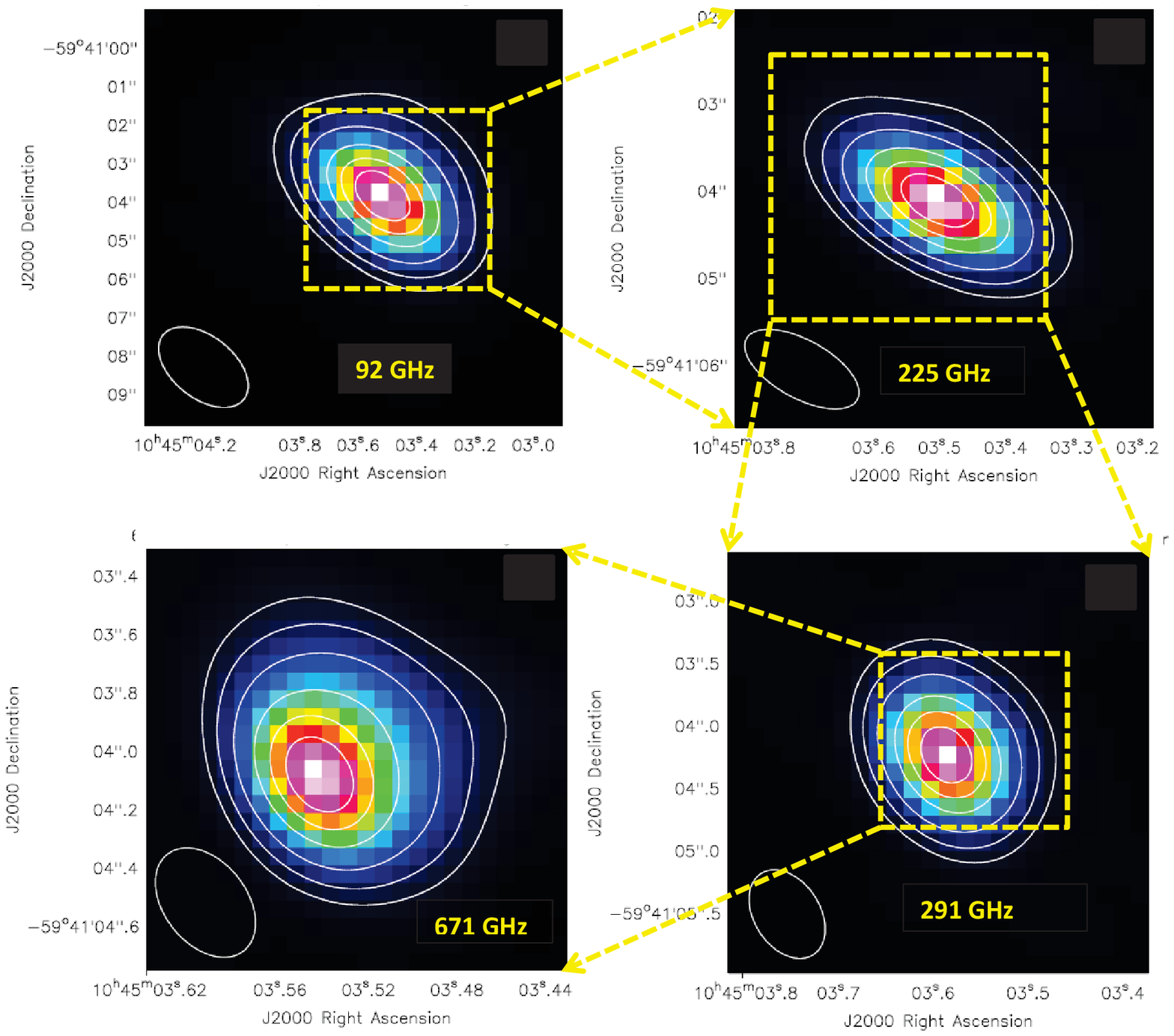}
\caption Contour maps of the continuum around $\eta$ Carinae. The peak of each map corresponds to the position of $\eta$ Carinae. The squares in the colored raster map indicate the cell size adopted for the cleaning algorithm. The  frequencies as well as the beam sizes are indicated in each map. The contours are  0.05, 0.1, 0.2, 0.4, 0.6 and 0.8 of the maximum flux density of the continuum given in column 4 of Table 1. The yellow boxes indicate the relative sizes of the maps at the different frquencies.
\end{center}
\end{figure*}

\section{Observations}

The observations were made with ALMA  (Atacama Large Millimetric/Submillimeter Array) as part of Early Science Cycle 0, on 2012 November 4 and 5, in four bands,  centered at 92 GHz (3.3 mm), 225 GHz (1.3 mm), 291 GHz (1.0 mm)  and 672 GHz (447 $\micron$)\footnote{This paper makes use of the following ALMA data: 2011.0.00497.S. ALMA is a partnership of ESO (representing its member states), NSF (USA) and NINS (Japan), together with NRC (Canada) and NSC and ASIAA (Taiwan), in cooperation with the Republic of Chile. The joint ALMA Observatory is operated by ESO, AUI/NRAO and NAOJ}. The 7.5 GHz bandwidth of the correlator was divided in four spectral windows, with resolution of 488 kHz, which corresponds to 1.5, 0.65, 0.5 and 0.22 km s$^{-1}$  for each frequency band, respectively. This configuration allowed the observation of continuum emission as well as the recombination lines H42$\alpha$ (85.7 GHz), H40$\alpha$ (99.0 GHz), H50$\beta$ (99.2 GHz) H30$\alpha$ (231.9 GHz), H28$\alpha$ (284.3 GHz) and H21$\alpha$ (662.4 GHz).  The 
corresponding He$\alpha$ lines are separated from the H$\alpha$ lines  by  $\Delta\nu/\nu = 0.000407$, or -122 km s$^{-1}$, and were also present in the spectra. Twenty three antennas were used in the observations, with a maximum baseline of 375 m. The source J1107-448 was observed for bandpass calibration and J1038-5311 and Titan for phase calibration and the latter also for flux calibration. The integration time for $\eta$ Carinae was 6 sec for the lowest frequency bands and 18 sec for the 672 GHz band.

We used the CLEAN algorithm present in CASA (Common Astronomy Software Applications) to construct the images.
The cell sizes used were $0\farcs45$, $0\farcs17$, $0\farcs13$ and $0\farcs05$ for the bands centered at 92, 225, 291 and 672 GHz, respectively.  Continuum maps were built excluding the channels in which line emission was present. Once these maps were obtained, the channels with line emission were analyzed, after subtracting the continuum contribution from the UV data. Line map cubes were obtained with velocity resolution of 2 km s$^{-1}$ for all lines, except for H21$\alpha$ and He21$\alpha$, for which the velocity resolution was 10 km s$^{-1}$ to compensate for the low signal to noise ratio of the individual channels. All velocities in this paper are referred to the local standard of rest (LSR).

 Besides the ALMA continuum observations of $\eta$ Carinae, we used 43 GHz data obtained with the Itapetinga\footnote{The Itapetinga radiotelescope is operated by the Brazilian Space Agency INPE (Instituto de Pesquisas Espaciais)} radiotelescope, located in S\~ao Paulo, Brazil, in October 30, 2012. The characteristics of the receiver and the observing mode are described in \citet{abr05}.

\begin{figure*}[!t]
\begin{center}
\includegraphics[width=15cm]{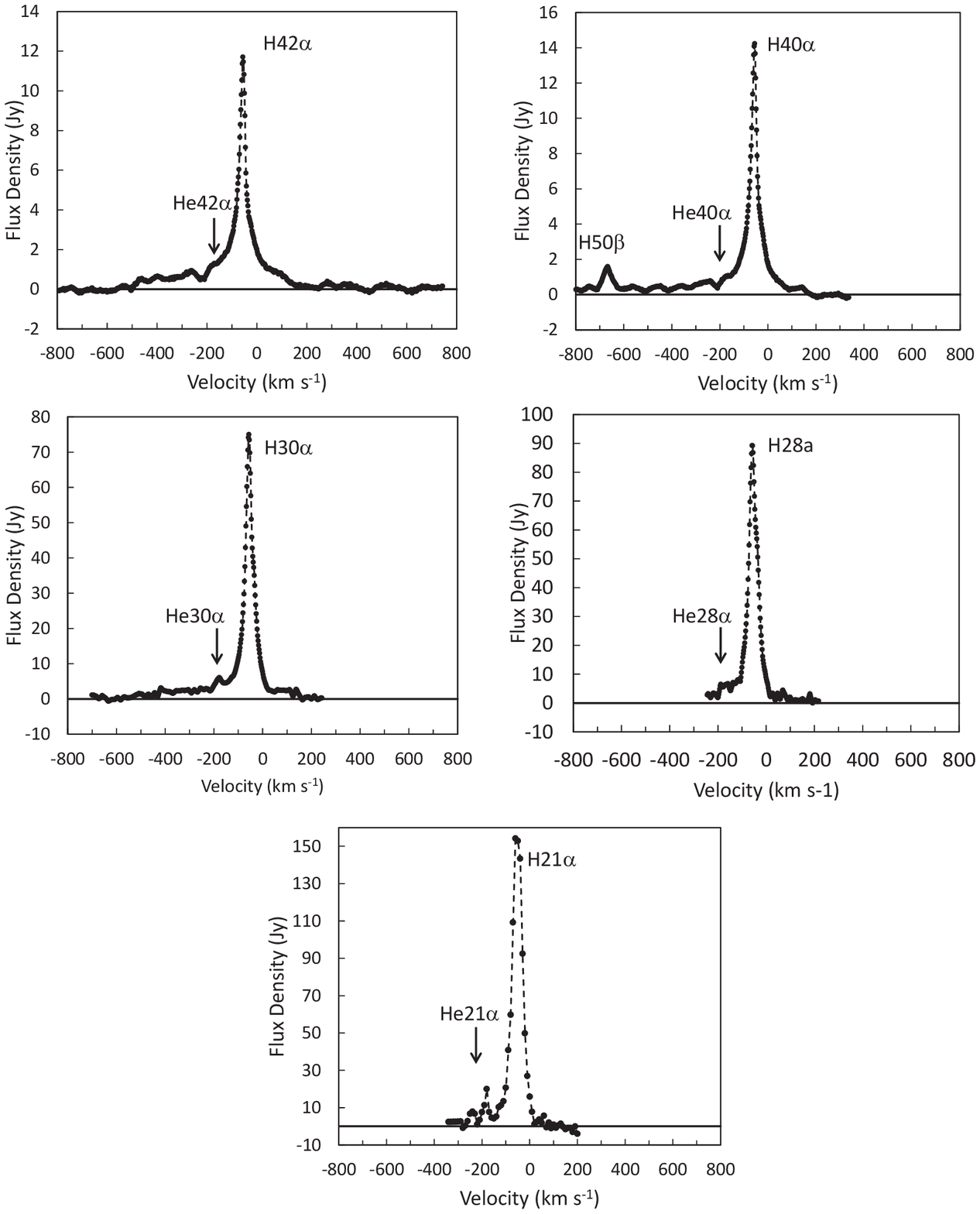}
\caption{H and He recombination line spectra integrated over the emission maps. The spectral resolution is 2 km s$^{-1}$ for all lines except the H21$\alpha$ and He21$\alpha$, for which the spectral resolution is 10 km s$^{-1}$}. 
\end{center}
\end{figure*}

\section{Results}

In Figure 1 we show the  continuum  maps of $\eta$ Carinae for the four observed bands. The  source remains unresolved at all frequencies.  In Table 1 we present, for each band, the central frequency, the size and orientation of the beam,  the flux density of the peak, in Jy/beam, the total flux density and its uncertainty.

Maps  in the radio recombination lines are similar to the continuum maps and are not shown in this paper.   The line profiles,  integrated over the whole  emission region are shown in Fig. 2. They include the lines H42$\alpha$, He42$\alpha$, H40$\alpha$, He40$\alpha$, H50$\beta$, H28$\alpha$, He28$\alpha$, H21$\alpha$ and He21$\alpha$. The frequencies were converted into velocities of the corresponding H$\alpha$ lines; in such configuration the He$\alpha$ lines appear separated by about -122 km s$^{-1}$ from the peak of the H$\alpha$ lines and the H50$\beta$ line by -613 km s$^{-1}$ from the H40$\alpha$ line.  The profiles were  superimposed   in Fig. 3, in the velocity range -300 km s$^{-1}$ to 100 km s$^{-1}$, to show the difference in flux density and width at half maximum between them. In Table 2 we present the frequency of the lines, the flux density and velocity of the peak and the line width. The central velocity (referred to the LSR) 
varies between -54 km s$^{-1}$ and -56 km s$^{-1}$, while the systemic velocity of the $\eta$ Carinae, derived from the velocity of the Homunculus lobes is -19.7 km s$^{-1}$ \citep{smi04}. The spectra clearly extends, with low intensity, from +100 km$^{-1}$ to -500 km s$^{-1}$.  The line width varies between 29 km s$^{-1}$ for the H42$\alpha$ line to 40 km s$^{-1}$ for the H21$\alpha$ line, larger than the expected thermal velocity.   

\begin{table}
\begin{center}
\caption{ Continuum parameters \label{tbl-1}}
\begin{tabular}{ccccc}
\tableline\tableline
 Freq. & beam size &  PA & Max & Flux   \\
 GHz   & arc sec   &  deg. & Jy/beam & Jy     \\
\tableline
 43.0  &  $144\times 144$ &     &       &  $8.7 \pm 1.0$ \\
 92.5  &  $2.88\times 1.83$ &  54 &  20.9 & $28.4 \pm 0.03$ \\
225.4  &  $1.52\times 0.75$ &  62 &  26.2 & $41.2 \pm 0.06$ \\
291.2  &  $0.83\times 0.60$ &  35 &  25.7 & $44.3 \pm 0.10$ \\
671.7  &  $0.45\times 0.32$&   42 &  22.1 & $40.2 \pm 0.09$ \\
\tableline
\end{tabular}
\end{center}
\end{table}

\begin{table}
\begin{center}
\caption{ H Line parameters \label{tbl-2}}
\begin{tabular}{crccc}
\tableline\tableline
Line & Freq. & Resolution &  Max. Flux & width  \\
   &  GHz  & km s$^{-1}$ & Jy &  km s$^{-1}$    \\
\tableline
H42$\alpha$  &  85.69 &  2 &  12 & 29 \\
H40$\alpha$  &  99.02 &  2 &  14 & 30 \\
H30$\alpha$  & 231.92 &  2 &  75 & 36 \\
H28$\alpha$  & 285.25 &  2 &  89 & 42 \\
H21$\alpha$  & 662.40 & 10 & 150 & 40 \\
\tableline
\end{tabular}
\end{center}
\end{table}

\begin{figure}[!t]
\begin{center}
\includegraphics[width=8cm]{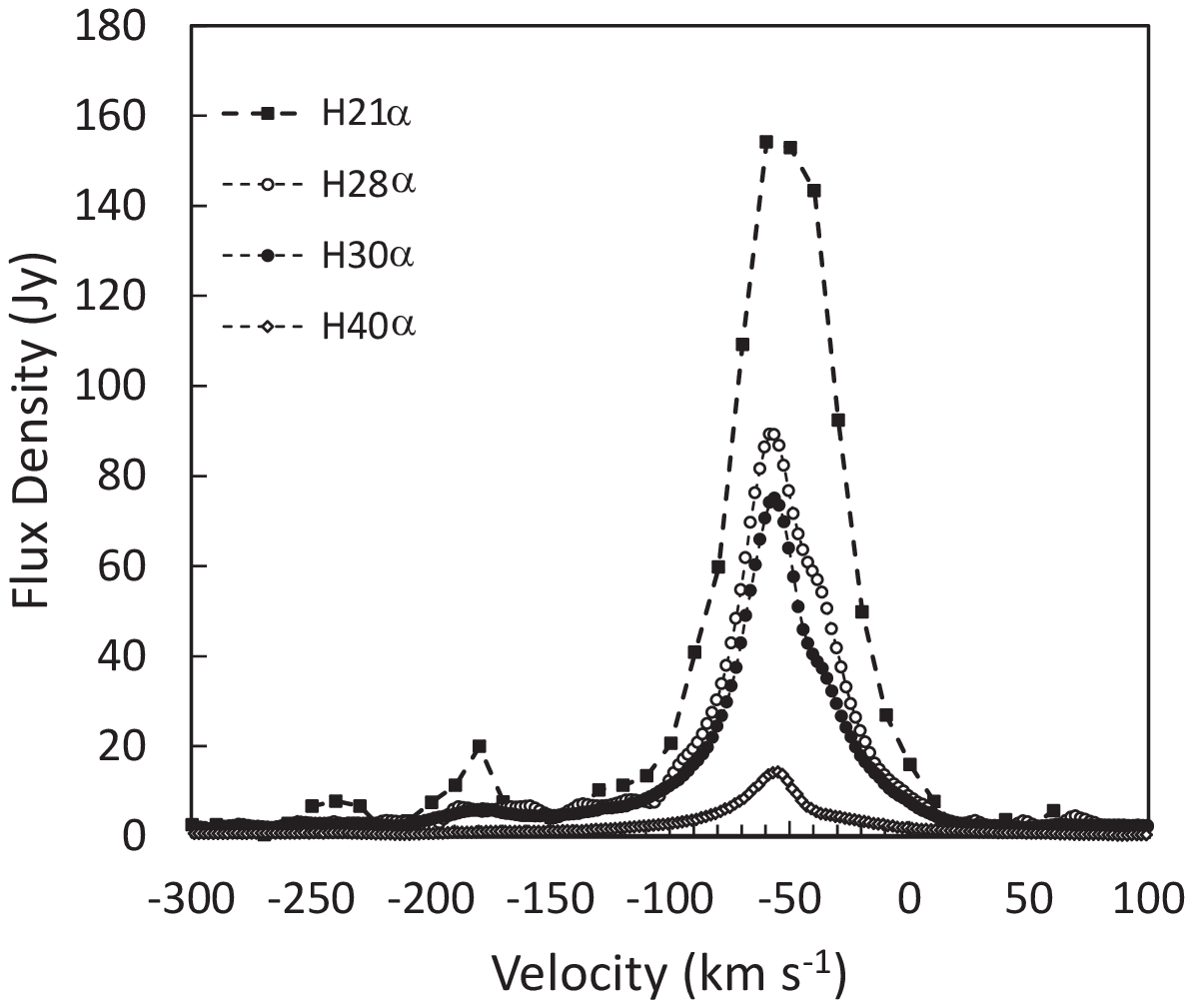}
\caption{Superimposed H and He recombination line spectra (40$\alpha$, 30$\alpha$, 28$\alpha$ and 21$\alpha$) integrated over the whole emission maps} 
\end{center}
\end{figure}

\begin{figure}[!t]
\begin{center}
\includegraphics[width=7cm]{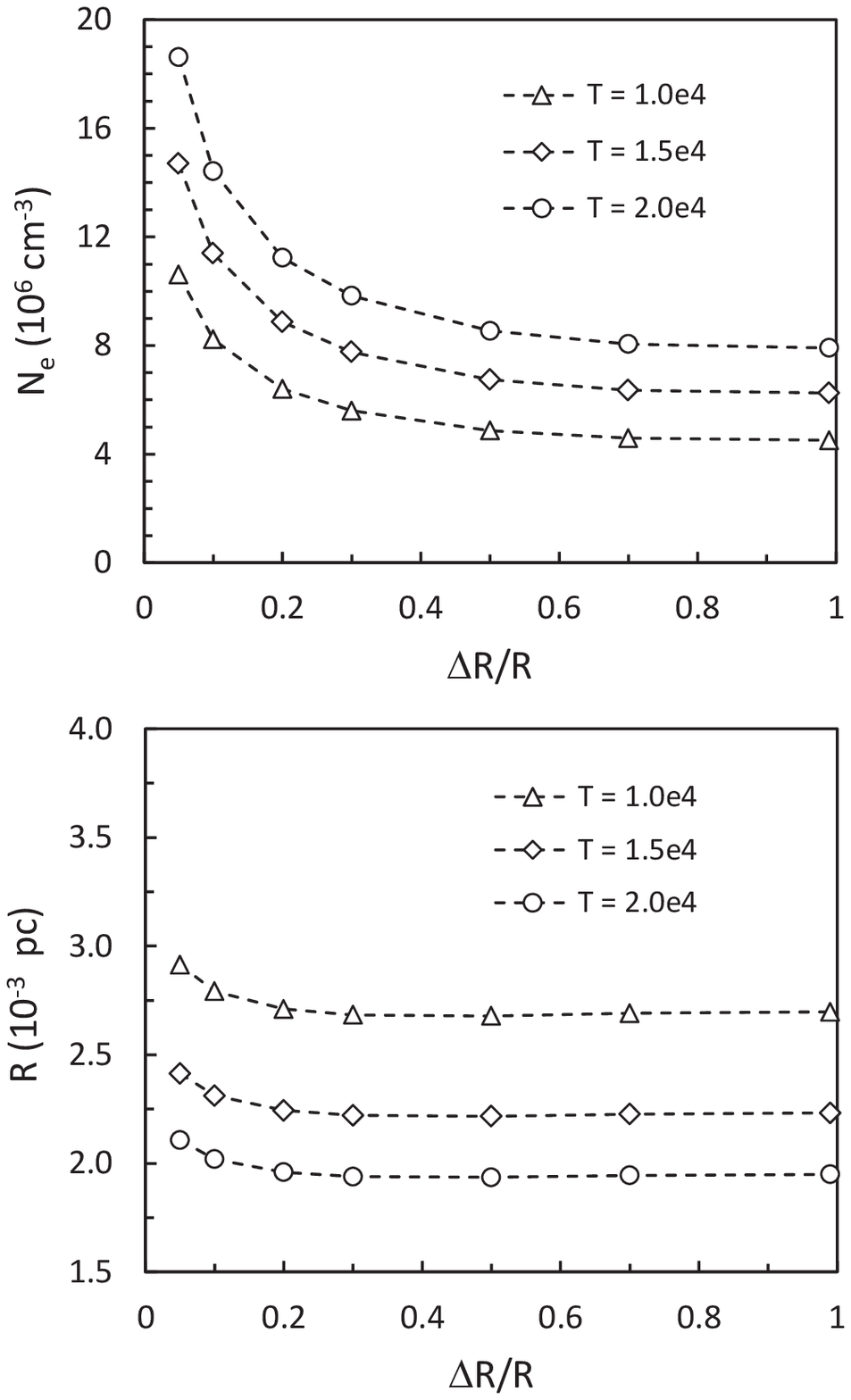}
\caption{Electron density $N_e$ (top) and radius of the emitting region $R$ (bottom) as a function of the relative shell size $\Delta R/R$ for all the models that fit the continuum spectrum for three values of the electron temperature $T_e$: $10^4$ K, $1.5\times 10^4$ K and $2\times 10^4$ K}.
\end{center}
\end{figure}

\begin{figure}[!t]
\begin{center}
\includegraphics[width=8cm]{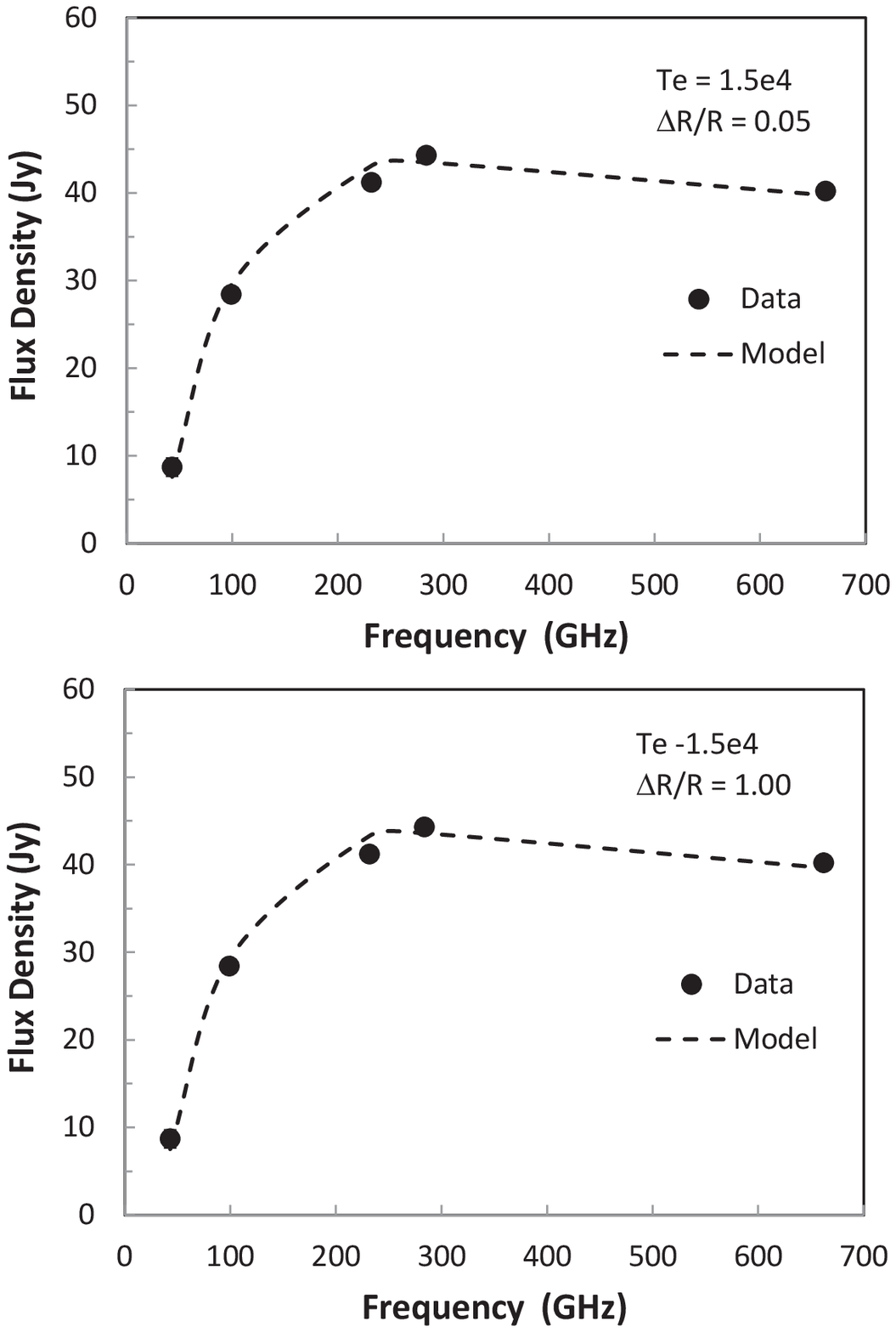}
\caption{Observed continuum flux density (points) and models (broken lines), for an  electron temperature of $1.5 \times 10^4$ K, the corresponding electron density and radius given in fig. 4, and two values of the shell width $\Delta R/R$: 0.05 (top) and 1.0 (bottom)}. 
\end{center}
\end{figure}

\begin{figure}[!t]
\begin{center}
\includegraphics[width=8cm]{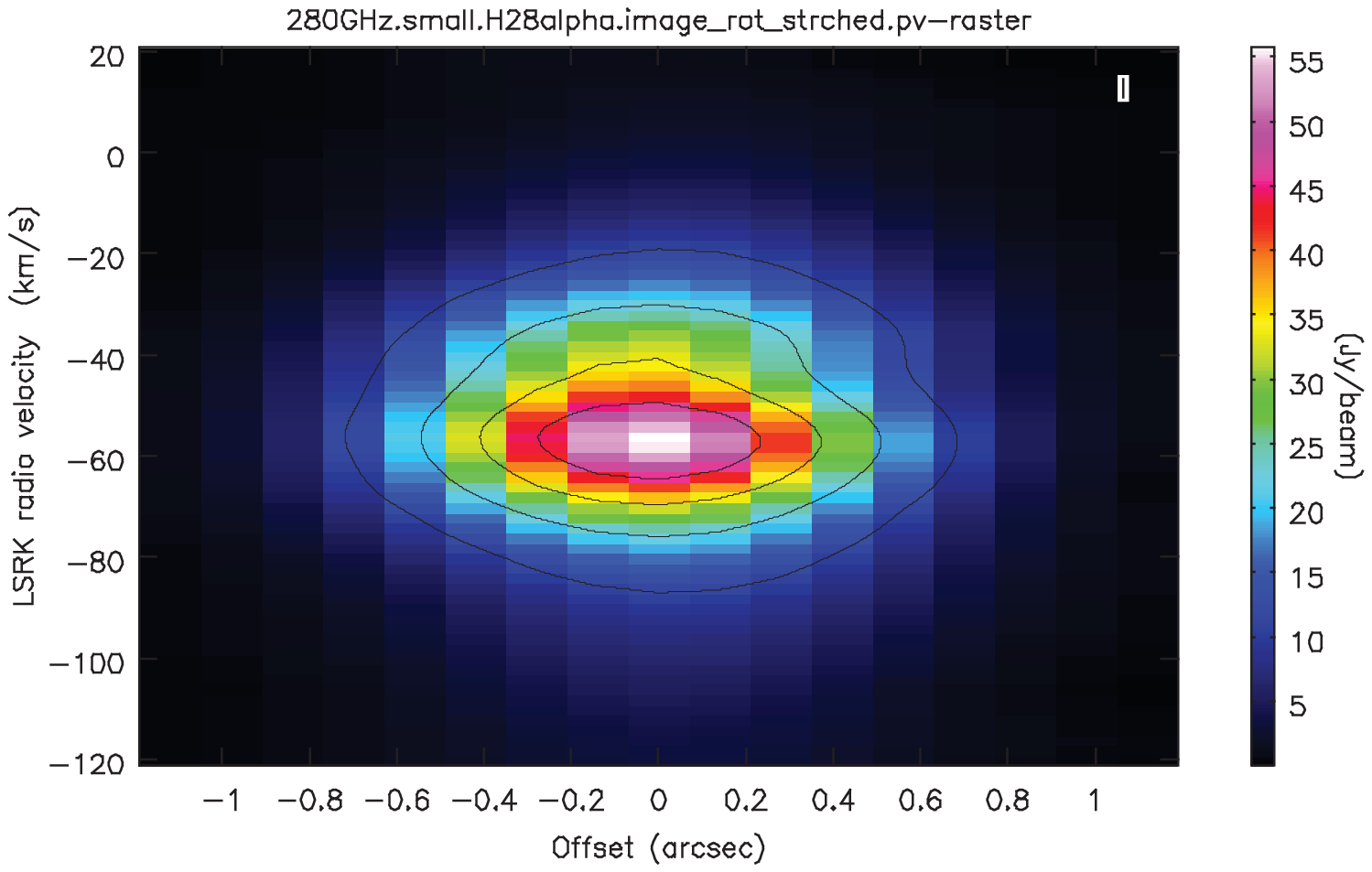}
\caption{Position velocity diagram for the H$28\alpha$ line, in the direction of the clean beam major axis, integrated by 5 pixels in the direction of the minor axis.  The contours are 0.2, 0.4, 0.6, and 0.8 of the maximum flux density, which is 55.9 Jy/beam}. 
\end{center}
\end{figure}

\section{Discusion}
\subsection{The continuum spectrum}
As shown in Table 1, the continuum flux density of   $\eta$ Carinae increases with frequency  up to 291 GHz and slowly decreases afterwards.  This spectrum in typical of bremsstrahlung from  compact \ion{H}{2} regions.
Since the source  is not resolved in our observations even at 670 GHz, we assumed  spherical symmetry and solved the radiative transfer equation for a shell of radius $R$, width $\Delta R$, and constant density and temperature. We added to the four frequencies observed by ALMA, a 43 GHz (7 mm) observation obtained at the same epoch with the Itapetinga radiotelescope ($2\farcm 4$ resolution), which gave a flux density of $8.7\pm 1$ Jy. 

\begin{figure}[!t]
\begin{center}
\includegraphics[width=6.9cm]{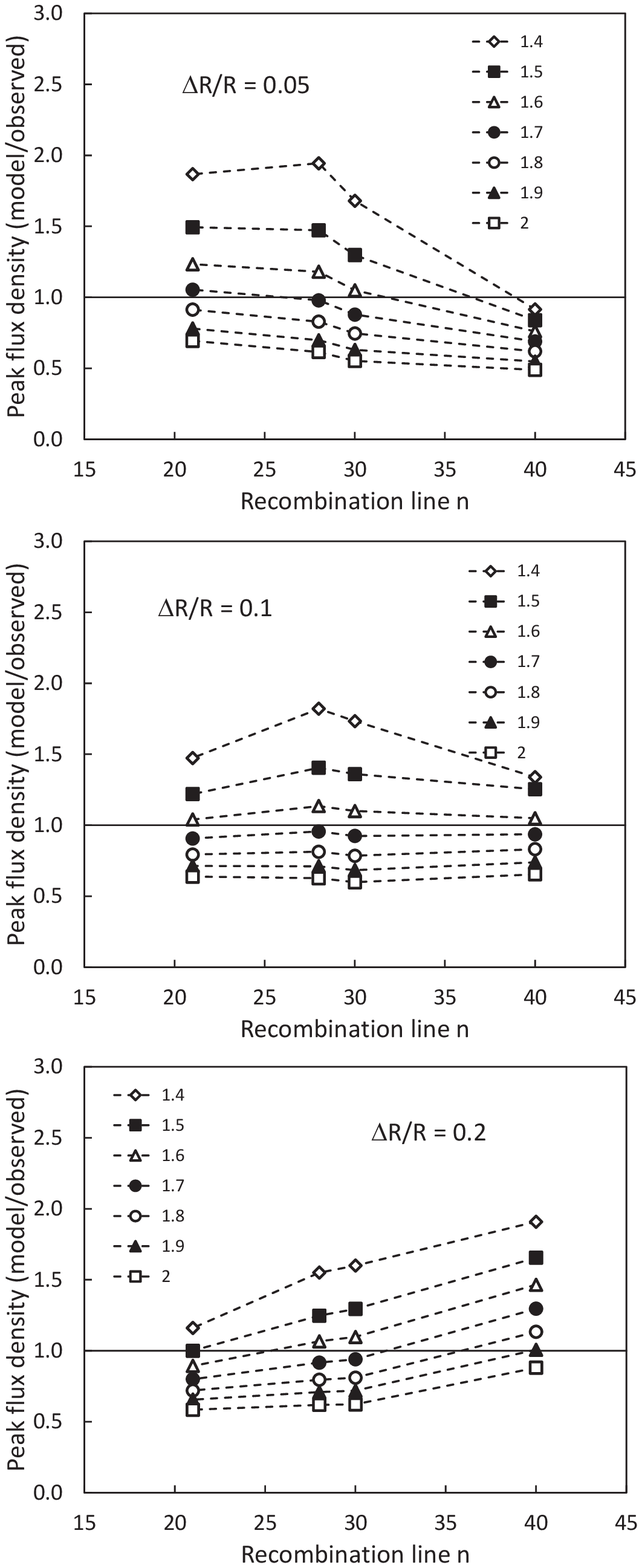}
\caption{Ratio of model and observed peak flux densities for the recombination lines H$40\alpha$, H$30\alpha$, H$28\alpha$ and H$21\alpha$ for electron temperatures ranging from $1.4\times 10^4$ K to $2.0\times 10^4$ K, and three values of the shell width $\Delta R/R:$ 0.05, 0.1 and 0.2 from top to bottom, respectively}
\end{center}
\end{figure}
 To calculate the flux density  at each frequency band, we divided the source in rectangular cells 
of size $(\Delta x, \Delta y)$ in the plane of the sky and solved the radiation transfer equation along the line of sight ($z$ coordinate) in each cell, using the continuum free-free emisivity equation (see Appendix A). We assumed that H is completely ionized and He is in the form of He$^+$, implying that $N_i = N_e$. We integrated   the flux density of each cell  $S_\nu^c(x,y)$ over the emitting region using as parameters the relative shell width $\Delta R/R$ and the electron temperature $T_e$, and determined the electron density and radius necessary to fit the observed spectrum.  Good fits were obtained for all combinations of electron temperature and shell width; in Fig. 4 we show the electron density and radius vs. the relative shell widths $\Delta R/R$ for three values of $T_e$: $10^4$, $1.5\times 10^4$  and $2\times 10^4$ K. We can see that the radius of the emitting region varied between a maximum value of $2.9\times 10^{-3}$ pc  for $T_e=10^4$ K and $\Delta R/R =0.05$ and a minimum of $1.9\times 10^{-3}$ pc for $T=2\times 10^4$ K and $\Delta R/R=1$. These dimensions correspond, at the distance of $\eta$ Carinae, to $0\farcs26$ and $0\farcs17$ respectively, both smaller than the resolution of our observation.
Two examples of continuum spectrum fittings are presented in Fig. 5, they correspond to $T_e = 1.5\times 10^4$ K and $\Delta R/R =$ 0.1 and 1, respectively.

\subsection{Recombination line emission}
To solve the transfer equation for line emission it is necessary to take into account other parameters, namely the thermal, turbulent and bulk velocities, as well as the line profiles  (detailed description of the calculations is given in appendix B).

To determine the  velocity distribution of the different cells from the observations, we constructed the position-velocity diagram for all the H lines, in the direction of the clean beam major axis, integrated  by 5 pixels in the direction of the minor axis. In Fig. 6 we show as an example, the diagram for the H28$\alpha$ line. We can see that the velocity distribution is compatible with that of a plasma cloud that has a velocity dispersion of about 40 km s$^{-1}$ and  is moving with a bulk velocity  of about -56 km s$^{-1}$.  

\begin{figure*}[!t]
\begin{center}
\includegraphics[width=17cm]{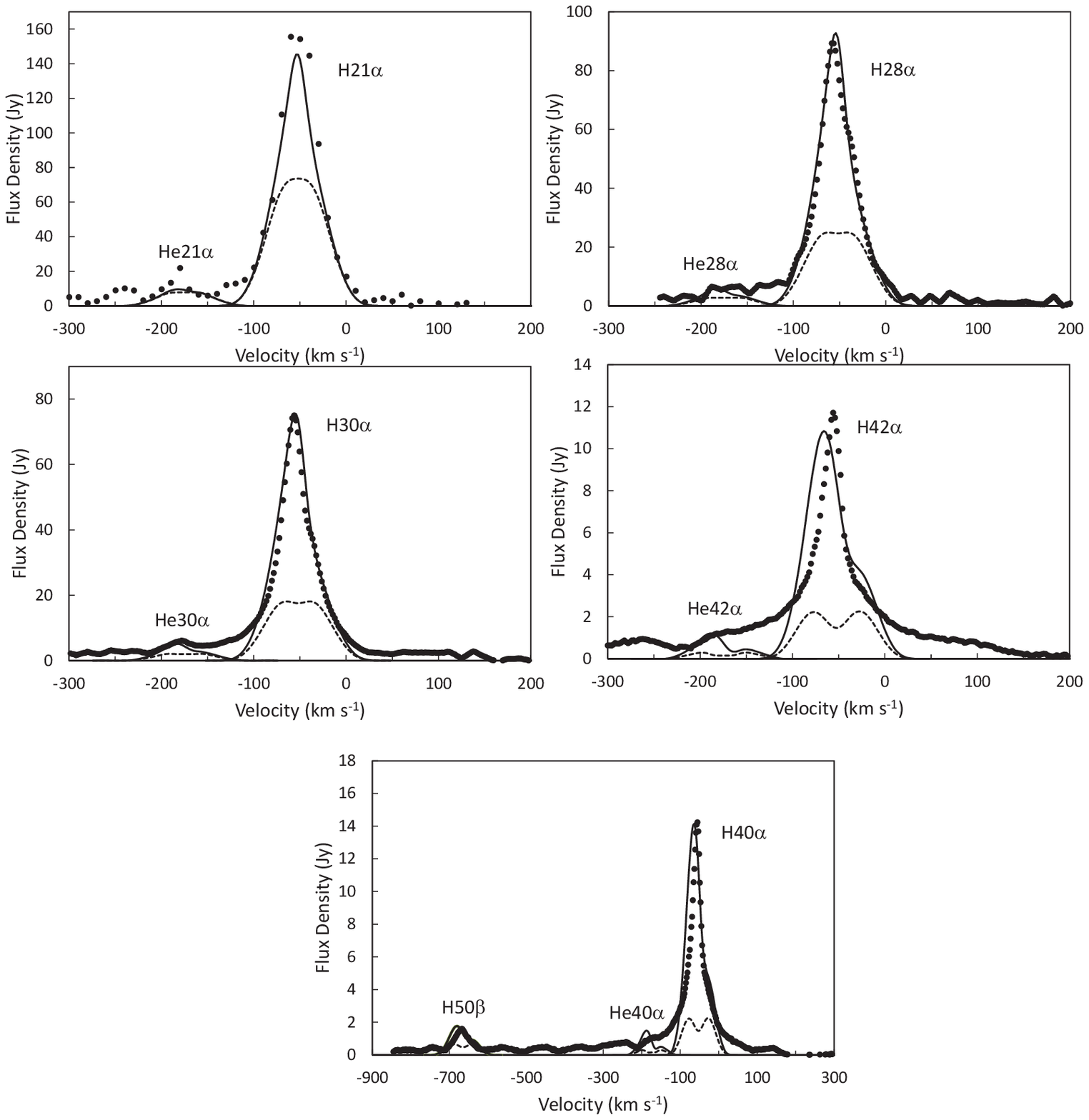}
\caption{Observed (points) and  best model (solid line) line profiles for the H$\alpha$ and He$\alpha$ recombination lines, as well as for the H$50\beta$ line. The dashed line represents the line profiles obtained assuming LTE conditions.} 
\end{center}
\end{figure*}

With this information we solved  the transfer equation (B1), and integrated the resulting flux density over the emitting region, using the combination of $N_e$ and $R$ obtained from the fitting of the continuum spectrum, for several combinations of $T_e$ and $\Delta R/R$. We were not able to reproduce the line profiles adding only turbulence to the thermal velocity, for that reason we assumed $T_D = T_e$ in equation (B6) and introduced an expansion velocity, with a linear gradient across the shell. We obtained the best fitting for an expansion velocity of 60 km s$^{-1}$ in the internal part of the shell and of 20 km s$^{-1}$ in the external part. To obtain the physical parameters of the emitting region we calculated the ratio of the model and observed peak flux densities for  the recombination lines H40$\alpha$, H30$\alpha$, H28$\alpha$ and H21$\alpha$, for values of  $T_e$ ranging from $1.4\times 10^4$ K to $2\times 10^4$ K and for three values of $\Delta R/R$: 0.05, 0.1 and 0.2. The results are presented in Fig. 7, where we can 
see that the ratio becomes close to unity for all lines for $T_e \sim 1.7\times 10^4$  and $\Delta R/R \sim 0.1$. 

In Fig. 8 we present the observed and model line profiles  for $T=1.68\times 10^4$ K and a relative shell width $\Delta R/R = 0.1$. The other parameters of the fitting are $N_e = 1.25 \times 10^7$ cm$^{-3}$ and $R= 2.2 \times 10^{-3}$ pc, values that also fit the continuum spectrum. We also show the observed and model line profiles for  H50$\beta$ and  for all the  He recombination lines, as well as all the H line profiles calculated assuming LTE. In all cases the bulk velocity of the model is $-52$ km s$^{-1}$.

We can see that the model reproduces very well the high frequency recombination line profiles (H21$\alpha$, H28$\alpha$ and H30$\alpha$). The fitting is not so good for the H40$\alpha$ and H42$\alpha$ lines, but we have to take into account that the continuum is optically thick at these frequencies, and in NLTE conditions the line amplification, being proportional to $\tau_{\nu}^l+\tau_{\nu}^c <0$ is very sensitive to the value of $\tau_\nu^c >1$. Therefore, variations of the electron density across the shell or departure from spherical symmetry in the emitting region will affect more the low frequency than the high frequency lines, for which $\tau_\nu^c <<1$. 
Besides, low intensity line emission is seen in the H40$\alpha$ and H42$\alpha$ lines extending up to velocities of $-500$ km s$^{-1}$, they must probably be produced in the Little Homunculus and can absorb part of the low velocity line emission.
 In the case of departure from spherical symmetry, the velocity distribution necessary to fit the line profiles will depend on the density and temperature distributions and  the solution will be degenerate. A more realistic model could be developed once data with finer angular resolution becomes available.

\subsection{The "Baby Homunculus"}

In the previous section we showed that the continuum spectrum and recombination line profiles of $\eta$ Carinae can originate in a spherical ionized shell of radius $R = 6.6 \times 10^{15}$ cm and width $0.1 R$, equivalent to $0\farcs2$ and $0\farcs02$ in the plane of the sky, respectively. The shell has constant density $N_e=1.27\times 10^7$ cm$^{-3}$ and temperature $T_e=1.7\times 10^4$ K, and it is moving with a bulk velocity of $-52$ km$s^{-1}$ in the LSR velocity frame, and expanding with a velocity that varies linearly across the shell between 60 km s$^{-1}$ in the inner border of the shell and 20 km s$^{-1}$ in the outer border. This velocity gradient, together with the high gas temperature suggests the existence of a slow shock, produced by the interaction of the expanding shell and the surrounding media. However, the ionizing source must be the secondary star, since the radio light curves present the same 5.5y periodic behavior observed at optical wavelengths and at  X-rays \citep{cox95a,abr05}. The 
observed He recombination lines indicate that the ionizing star must have a surface temperature higher than that of $\eta$ Carinae \citep{meh10}.

This shell was probably ejected by $\eta$ Carinae in an episode similar to those which formed the Homunculus and the Little Homunculus, and for that reason we called it the "Baby Homunculus". The bulk velocity of the shell may reflect the orbital velocity of $\eta$ Carinae in the direction of the line of sight at the  corresponding orbital phase. The ejection epoch is difficult to determine because, although we know the size of the shell, our model assumes a gradient in the expansion velocity. However, looking at the historical light curve at optical wavelengths compiled by \citet{fer09}, shown in fig. 9, we can guess that the ejection occurred around 1941, when there was another sudden increase in stellar luminosity, similar that which formed the Little Homunculus. This would require a mean expansion velocity of about 30 km s$^{-1}$ to attain  the present size of 0.0022 pc. 

\subsection{Dynamical evoloution of a wind blown shell}

The origin of Eta Carinae massive bursts is not clear yet, but may be possibly related to the S Dor 
cycles of luminous hot stars, though much more massive and luminous than those of other 
luminous blue variable (LBV) stars \citep{pasto2010}. LBV variability is categorized 
by its change in brightness \citep{hd94}: i) {\it giant eruptions}, with brightness 
changes of $\Delta m > 2$ in timescales of few years, ii) {\it eruptions}, with $\Delta m < 2$ in 
timescales of few months up to a couple of years, and iii) {\it small oscillations}, with 
$\Delta m < 0.5$, in timescales of few days to few weeks. 

Based on the observed light curve of Eta Carinae shown in Fig. 9, the 1940's burst 
resulted in $\Delta m_V \simeq 1$ and may be related to a LBV eruption. Similar case 
might be that of AG Car during its 1990's eruption, in which its brightness increased by $\Delta m \sim 2$
 during a period of almost 
3 years. During the maximum, the star 
reached a mass loss rate of $\dot{M} \sim 10^{-3.8}$M$_\odot$yr$^{-1}$, 
almost one order of magnitude larger than that at the minimmum, though with reduced 
wind speed of a few hundred of km s$^{-1}$ \citep{stahl01}. 

 A similar case migh be an increase in the continuous mass loss rate and decrease in the wind velocity  of $\eta$ Carinae; once the 
the wind mass loss rate  decreased, the wind  terminal speed increased again. This two-phased
wind evolution could be responsible for the formation of the shell-like structure that expands outward 
relative to the central object. The internal part would be a low density expanding cavity created 
by the fast and less massive post-eruption wind, surrounded by a denser shell 
expanding at lower speeds, a relic of the massive wind generated during the eruption.

The dynamical evolution of an expanding shell driven by a continuous stellar wind
is well described by the classical work of \citet{weaver77}. 
According to that work, the radius $R_1(t)$ of the expanding shock formed between the wind, of velocity $v_w$ and mass loss rate $\dot{M}$, and its surrounding medium of density $\rho_0$ is:

\begin{equation}
R_1(t) \simeq L_w^{1/5} \rho_0^{-1/5} t^{3/5},
\label{eq:weaver}
\end{equation}

\noindent
while the expanding velocity $v_1(t)$ would be given by:

\begin{equation}
v_1(t) \equiv \dot{R} \simeq \frac{3}{5} L_w^{1/5} \rho_0^{-1/5} t^{-2/5},
\label{eq:weaver2}
\end{equation}

\noindent
where $L_w \simeq \dot{M} v_w^2$ is the wind luminosity and $t$ is the elapsed time from the start of the
eruption. 
Once the eruption is over, and the star goes back to quiescent state at $t=t_{\rm trans}$, the ejecta will 
expand further but without any input of energy at its interior. At this stage, for $t>t_{\rm trans}$, 
Eq.\ref{eq:weaver} above is no longer valid. 
The dynamics of the expanding shell will then be better described as a blast wave. 
Here, two different phases are possible depending on the cooling efficiency of the shocked 
gas. The first is the Sedov solution for an adiabatically expanding shell, while the second 
is a momentum conserved snow-plow shell. The typical cooling timescale is given as 
$\tau_{\rm cool}\simeq kT/n\Lambda \sim 5 \times 10^{-3} n^{-1}$ yr, where $\Lambda$ is the energy loss rate, $T \sim 10^5$ K, and the number density $n$ is given in cm$^{-3}$ . 
In the case of $\eta$ Carinae, the whole process occurs in high density regimes (the stellar wind and surrounding medium), 
i.e. $n > 10^3$ cm$^{-3}$. Therefore we may consider the expansion of the post-eruption shell 
in the radiative (snow plow) regime, as:

\begin{equation}
R_2(t) = R_{\rm trans} \left(\frac{t}{t_{\rm trans}}\right)^{1/4},
\label{eq:spr}
\end{equation}

\noindent
and

\begin{equation}
v_2(t) = v_{\rm trans} \left(\frac{t}{t_{\rm trans}}\right)^{-3/4},
\label{eq:spv}
\end{equation}

\noindent
where $v_{\rm trans}$ stands for the velocity at the transition time  
which can be obtained from equations \ref{eq:weaver} and \ref{eq:weaver2}.

The  mass of the shell $M_{\rm shell}$, gives an upper limit to the mass ejected by the star:
\begin{equation}
M_{\rm shell}=\mu m_H N_e\pi R_2^2(t_2)\Delta R_2=\dot{M}t_{\rm trans},
\label{eq:mshell}
\end{equation}

\noindent
where $t_2$ is the time elapsed since the beginning of the ejection until the observation; it was taken as 71 yr,  linking the beginning of the ejection to the luminosity increase observed in 1941.

Using equations \ref{eq:weaver}-\ref{eq:mshell} and our shell model parameters we obtain:

\begin{equation}
t_{\rm trans}=\frac{0.002}{\dot M},
\end{equation}

\begin{equation}
v_w^2 = 0.18 \dot{M}^{3/4}N_0,
\end{equation}

\noindent
and
\begin{equation}
v_2(t_2)=\frac{3}{5}\frac{R_2(t_2)}{t_2},
\label{eq:v2t2}
\end{equation}

\noindent
where $\dot{M}$ is given in $M_{\odot} y^{-1}$, $t_{\rm trans}$ in years, $N_0$ in cm$^{-3}$ and $v_w$ in km s$^{-1}$.

From eq. \ref{eq:v2t2} we immediately obtain $v_2(t_2)= 19$ km s$^{-1}$ in excellent agreement with the expansion velocity derived for the shell. The wind velocity during the eruption and its duration vs. the mass loss rate are presented in Fig. 10 for two values of the ambient density: $10^5$ and $10^6$ cm$^{-3}$. It is clear that the wind velocities and mass loss rates are compatible with what is expected in an LBV eruption.

In spite of the fact that we cannot strictly constrain the wind 
parameters during the eruption,  the model used to fit the 
observed data gives reasonable values, compared 
to those expected for LBV eruptions, i.e. wind velocities of 
$v_w = 50 - 200$ km s$^{-1}$, and mass loss rates of 
$\dot{M}=  10^{-2.5}-10^{-1.5} $M$_\odot$yr$^{-1}$ $\sim 10 - 100$ times 
larger than the current value estimated for 
$\eta$ Carinae, lasting between one year and one month, respectively.

\subsection{The high excitation optical lines and the  Weigelt Blobs}

The densities and temperatures of the shell model predict also the existence of high excitation optical emission lines, which were in fact observed in the $0\farcs1-0\farcs3$ surroundings of $\eta$ Carinae \citep{meh10}. Spatial maps of these lines show the presence of compact condensations, known as the Weigelt Blobs, which  present lines of  highly ionized ions like [NeIII] and [FeIII]; their presence require densities $N_e\geq 10^7$ cm$^{-3}$ and temperatures $T_e\sim 2\times 10^4$ K. Their line of sight velocities are small relative to the wind velocity ($-47$ km s$^{-1}$ in the geocentric reference frame or $-58$ km s$^{-1}$ in the LSR), but they were never related to the source of the radio recombination lines observed by \citet{cox95b}.

Although these condensations are known since 1986 \citep{wei86}, their nature remained a puzzle.  As a consequence of our study, it is possible  that the Weigelt blobs are part of the Baby Homunculus; not only their velocities coincide, but also their densities and temperatures. 

The behavior of the radio emission and the optical light curve of the Weigelt Blobs along the different phases of the binary cycle are also similar.  In fact, \citet{meh10}  reported that the optical light curve of the high-excitation lines increased up to the middle of the binary phase between 1988 and 2003.5, presenting another maximum before the sharp minimum, which was also observed in the radio light curve \citep{abr05}. This behavior can be understood if the shell is ionized by the secondary star. During the minimum, the ionizing photons do not reach the high density shell and the plasma recombines with time scales of days. When the ionizing photons reach the Baby Homunculus again, they ionize the gas, but the ionizing front propagates  slower than the recombination  time scale, explaining the gradual increase in the observed flux density. 

   Other characteristics of the Weigelt blobs can be explained if they are part of the expanding shell, like the 10 km s$^{-1}$ difference in velocity between the high-ionization lines  and those of the low-excitation and low-ionization levels  found by \citet{smi04} that can be a consequence of the velocity gradient across the shell. 

The ejection time of the Weigelt blobs is also controversial, ranging from 1890 \citep{smi04} to 1941 \citep{dor04}, the last one compatible with our interpretation that the shell was ejected in 1941.

\section{Conclusions}

 In this paper we presented millimeter and submillimeter observations of the continuum emission and H and He recombination lines of $\eta$ Carinae obtained with ALMA in November 2012.

The continuum spectrum is characteristic of thermal bremsstrahlung,  raising with frequency up to 230-280 GHz, frequency at which the source became optically thin, and decreasing slowly at higher frequencies. The high flux density observed in the recombination lines requires NLTE conditions and the presence of the He lines indicate that the ionizing star should have a surface temperature  higher than that attributed to $\eta$ Carinae.

We modeled the emitting region as a spherical shell, moving in the direction of the observer with a bulk velocity of $-52$ km s$^{-1}$ and expanding with velocities ranging between 20 and 60 km s$^{-1}$. These velocities were needed to reproduce the observed line profiles. The continuum spectrum and the intensity of the recombination lines could be better reproduced by a shell of radius $R=0.0022$ pc, width $\Delta R=0.1R$, electron temperature $T=1.7\times 10^4$ K and electron density $N_e=1.25\times 10^7$ cm$^{-3}$.
 The results of this model could be checked with future observations, when longer baselines and a larger number of ALMA antennae become available.

The physical parameters of the shell and the behavior of the radio light curve along the orbital phase are similar to those of the Weigelt blobs, and we suggest that these blobs are part of the shell.  We also suggest that the shell was ejected around 1941, when there was a sudden increase in the optical luminosity of $\eta$ Carinae.  As this increase is similar to that observed when  the Little Homunculus was formed, we proposed that the newly discovered shall be known as the 'Baby Homunculus'.

\acknowledgments

This work was supported by the Brazilian agencies CAPES, CNPq and FAPESP.
DFG thanks the European Research Council (ADG-2011 ECOGAL), and
Brazilian agencies CNPq (no. 300382/2008-1), CAPES (3400-13-1)
and FAPESP (no.2011/12909-8) for financial support.

\begin{figure}[!t]
\begin{center}
\includegraphics[width=7cm]{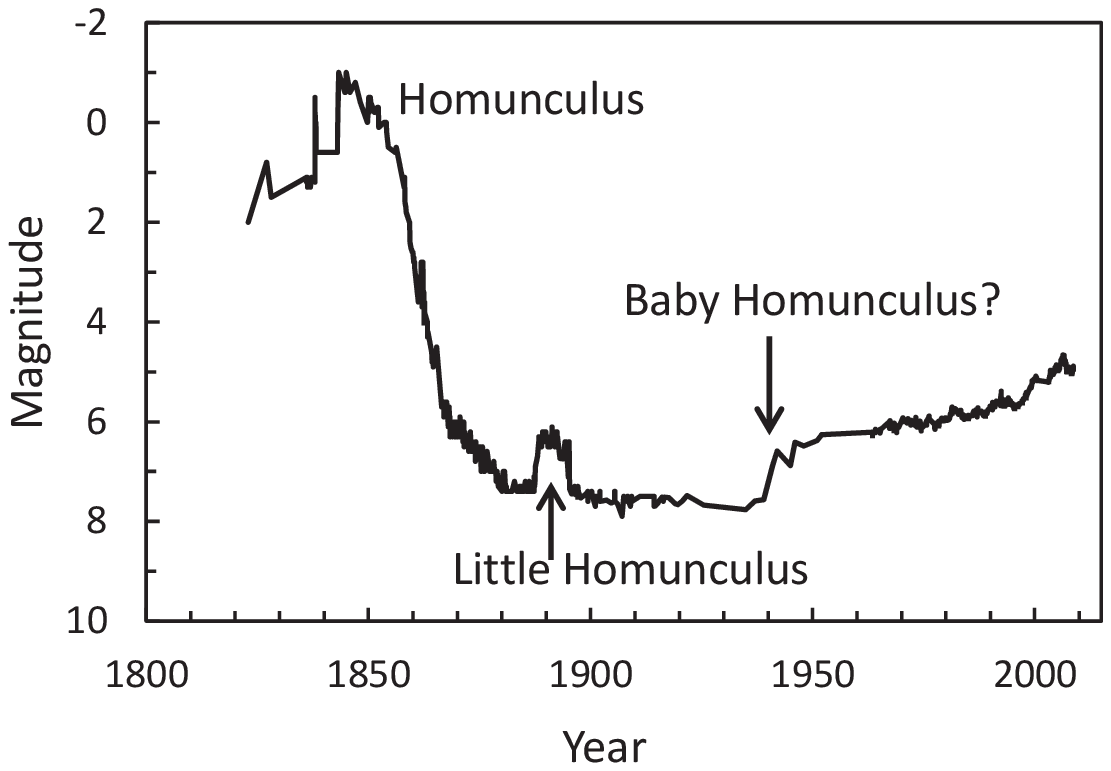}
\caption{Historical light curve of $\eta$ Carinae, built using data compiled by \citet{fer09}}. 
\end{center}
\end{figure}

\begin{figure}[!t]
\begin{center}
\includegraphics[width=7cm]{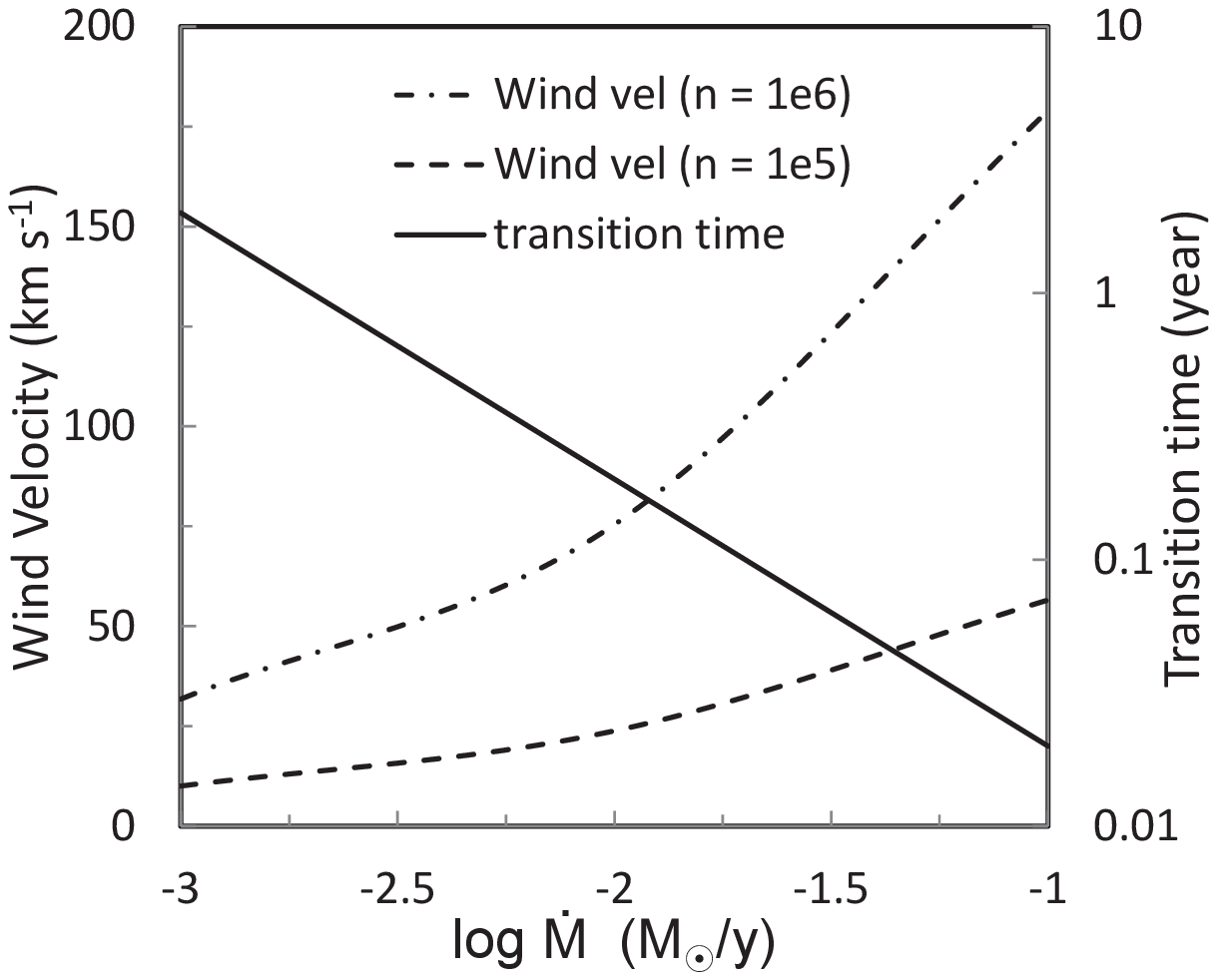}
\caption{Wind velocity (left axis) for two values of the ambient density: $10^5$ cm$^{-3}$ (dash line) and $10^6$ cm$^{-3}$ (dot-dash line), and duration of the BLV outburst (right axis, solid line) as a function of the mass loss rate during the outburst}. 
\end{center}
\end{figure}

\clearpage

\appendix

\section{Bremsstrahlung}

Let us consider an ionized source with spherical symmetry, formed by a shell of radius $R$, width $\Delta R$, and constant electron density  and temperature. 
The source is divided in rectangular cells of size $(\Delta x, \Delta y)$ in the plane of the sky and the radiation transfer equation is solved along the line of sight ($z$ coordinate) in each cell. 

The resulting continuum flux density $S_\nu^{\rm c}$ is given by:
\begin{equation}
S_\nu^c(x,y) = \frac{2\pi\nu^2}{c^2}kT_e[1-\exp(-\tau_\nu ^c(x,y))]\frac{\Delta x\Delta y}{D^2},
\end{equation}

\noindent
where $T_e$ is the electron temperature, $\nu$ the frequency, $k$ the Boltzman constant, $\tau_\nu^c(x,y)$ the optical depth for free-free absorption integrated along the line of sight, and $D$ the distance from the source to the observer.

Since we assumed constant electron temperature and density  in the emitting region, the optical depth can be written as $\tau_\nu^c(x,y) = \kappa_\nu^c \ell(x,y)$, where $\ell(x,y)$ is the depth of the absorbing region along the line of sight and the absorption coefficient for free-free radiation $\kappa_\nu^c$ is given by \citep{bro72}:

\begin{equation}
\kappa_\nu^c = 6.94\times 10^{-8}\frac{N_e N_i}{\nu^2}\left(\frac{10^4}{T_e}\right)^{3/2}g_{ff}~~ \rm pc^{-1},
\end{equation}

\noindent
where $N_e$ and $N_i$ are the electron and ion number densities and $g_{ff}$ is the Gaunt factor, given by:

\begin{equation}
g_{ff}= (4.69+1.5\ln T_e-\ln \nu) 
\end{equation}

\section{Recombination Lines}

 The total  (continuum plus line) flux density $S_\nu^{\rm l+c}(x,y)$ of each cell can be obtained from: 
\begin{eqnarray}
S_{\nu}^{l+c}(x,y) = \frac{2\pi\nu^2}{c^2}kT_e\frac{\Delta x\Delta y}{D^2} \nonumber \\
\times\int_0^{\ell(x,y)} (\kappa_\nu^c+b_{n+m}\kappa_{\nu,LTE}^l)\exp(-\tau_\nu^{l+c}) dz,
\end{eqnarray}
\noindent
where 
\begin{equation}
\tau_\nu^{l+c}(x,y,z)=\int_0^z (b_n\beta_{n, n+m}\kappa_{\nu,LTE}^l + \kappa_\nu^c) dz',
\end{equation}
\noindent
$\kappa_\nu^c$ is given by eq. A2, $\kappa_{\nu,LTE}^l$, the LTE line absorption coefficient  for the transition between levels with quantum numbers $n$ and $n+m$ is given by \citep{bro72}:
\begin{eqnarray} 
\kappa_{\nu,LTE}^l = 1.064\times10^{-12}\frac{N_eN_+}{\nu}\left(\frac{10^4}{T}\right)^{2.5}mK(m) \nonumber \\
  \times \exp(\chi_n/kT)(\nu\phi_{\nu}),
\end{eqnarray}
\noindent
where $N_+$ is the number of ions (H$^+$ or He$^+$), $mK(m)$ is related to  the oscillator strength \citep{men68}, with values of 0.1908 for $m=1$ and 0.05266 for $m=2$, $\chi_n$ is the energy of level $n$,  $b_n$ is the population of  level $n$ relative to its population in local thermodynamic equilibrium, calculated by \citet{sto95} for several values of $n$, $N_e$ and $T_e$,   
\begin{equation}
\beta_{n,n+m}=\frac {1-(b_{n+m}/b_n)\exp(-h\nu/kT)}{1-\exp(-h\nu/kT)}, 
\end{equation}
$\phi_{\nu}$ is the normalized line profile, which can be calculated from:
\begin{equation}
\phi_{\nu}=\int_{-\infty}^{\infty}\phi_{\nu-\nu'+\nu_0}^P\phi_{\nu'}^D d\nu',
\end{equation}
\noindent
where $\phi_{\nu}^D$ represents the Doppler broadening profile, given by \citep{bro71}:
\begin{equation}
\phi_{\nu}^D = \frac{\alpha}{\pi^{1/2}\nu_0}\exp\{-[\alpha(\nu-\nu_0)/\nu_0]^2\},
\end{equation}
\noindent
with $\alpha = [m_ic^2/(2kT_D]^{1/2}$, where $m_i$ is the mass of the ion and $T_D$ is the Doppler temperature, which includes the thermal and turbulent velocities; $\nu_0$ is the rest frequency of the transition.

$\phi_{\nu}^P$ represents the pressure broadening profile, which can be written as:
\begin{equation}
\phi_{\nu}^P = \frac{\delta}{\pi}\frac{1}{(\nu - \nu_0)^2+\delta^2},
\end{equation}
\noindent
with $\delta=1/2\pi\langle vQ\rangle N_e$, where $\langle vQ\rangle$ is the total inelastic cross-section averaged over the Maxwellian distribution \citep{bro71}.

The width of the thermal line can be calculated from:
\begin{equation}
(\Delta v)^D= 7.13\times 10^{-5}c\left[\frac{m_H}{m_i}\left(\frac{T_D}{10^4}\right)\right]^{1/2},
\end{equation}
\noindent
which corresponds to 21 km s$^{-1}$ for a $10^4$ K ionized H plasma, 

The width of the pressure broadened line can be calculated from \citep{bae13}:
\begin{eqnarray}
(\Delta v)^P= 8\times 10^{-10}\frac {n^5 N_ec}{\nu Z^2 T_e^{0.1}} & n\leq 30,
\end{eqnarray}

\begin{eqnarray}
(\Delta v)^P= 6.7\times 10^{-9}\frac {n^{4.6} N_e c}{\nu Z^2T_e^{0.1}}~~ 30< n< 100,
\end{eqnarray}
\noindent
which corresponds to 2.5 km s$^{-1}$ for $n=42$, $T_e=10^4$ K and $N_e=10^7$ cm$^{-3}$ and only $5
\times 10^{-3}$ km s$^{-1}$ for $n=21$, in both cases much smaller than the thermal width, and for that reason presure broadening was not included in the calculations.

\end{document}